\documentclass[preprint,prc,aps,showpacs,showkeys,superscriptaddress]{revtex4}
\usepackage{ae,aecompl}
\usepackage[T1]{fontenc}
\usepackage[latin9]{inputenc}
\usepackage{amsmath}
\usepackage{float}
\usepackage{amssymb}
\usepackage{graphicx}
\usepackage{multirow}
\usepackage{mathtools}

\makeatletter

\pdfpageheight\paperheight
\pdfpagewidth\paperwidth

\newcommand{\noyau}[3]{\prescript{#2}{#3}{\mathrm{#1}}}

\begin{document}

\title{Benchmarking GEANT4 nuclear models for hadron therapy with 95 MeV/nucleon carbon ions}

\author{J. Dudouet}
\author{D. Cussol}
\author{D. Durand}
\author{M. Labalme}
\affiliation{LPC Caen, ENSICAEN, Universit\'e de Caen, CNRS/IN2P3, Caen, France}

\date{\today}

\begin{abstract}

In carbon-therapy, the interaction of the incoming beam with human tissues may lead to the production of a large amount of nuclear fragments and secondary light particles. An accurate estimation of the biological dose on the tumor and the surrounding healthy tissues thus requires sophisticated simulation tools based on nuclear reaction models. The validity of such models requires intensive comparisons with as many sets of experimental data as possible. Up to now, a rather limited set of double differential carbon fragmentation cross sections have been measured in the energy range used in hadrontherapy (up to 400~MeV/nucleon). However, new data have been recently obtained at intermediate energy (95~MeV/nucleon). The aim of this work is to compare the reaction models embedded in the GEANT4 Monte Carlo toolkit with these new data. The strengths and weaknesses of each tested model, i.e. G4BinaryLightIonReaction, G4QMDReaction and INCL++, coupled to two different de-excitation models, i.e. the generalized evaporation model and the Fermi break-up are discussed.

\end{abstract}



\maketitle

\section{Introduction}\label{sec:Intro}

The use of carbon ions in oncology is motivated by some balistic and biological advantages. Carbon ions allow  to better target the tumor while preserving the surrounding healthy tissues. However, the physical dose deposition is affected by the inelastic processes of the ions along the penetration path in human tissues \cite{Schardt96,Matsufuji03}. For instance, the number of incident ions reaching the tumor (at the Bragg peak depth) is reduced up to 70\% for 400~MeV/nucleon $^{12}$C in tissue equivalent material \cite{Haettner06}. Carbon beam fragmentation in human body leads to the production of secondary lighter fragments with larger ranges and larger angular spreadings. Such fragments have also  a different biological efficiency, which is strongly correlated to the linear energy transfer (LET). These effects, due to carbon fragmentation, result in a complex spatial dose distribution, particularly on healthy tissues. The influence of the secondary particles production is the highest beyond the Bragg peak where only secondary particles contribute to the dose. 

In view of the previous remarks, to keep the benefits of carbon ions in hadrontherapy requires a very high accuracy on the dose deposition pattern ($\pm$3\% on the dose value and $\pm$2 mm spatial resolution \cite{SFPM}). In planning a tumor treatment, the nuclear reactions need to be correctly evaluated to compute the biological dose all along the beam path. Monte Carlo methods are probably the most powerful tools to take into account such effects. Even though they generally cannot be directly used in  clinic because of a too long processing time, they can be used to constrain and optimize analytical treatment planning system (TPS) \cite{Sihver09,Kramer10} or to generate complete and accurate data bases \cite{Toshito07,Golovchenko02,Schall96,Haettner06,Napoli12}.

The ability of Monte Carlo codes to reproduce differential yields of charged fragments from carbon fragmentation has been recently studied. B\"ohlen {\it et al.}~\cite{Bohlen10} studied the prediction capability of FLUKA~\cite{Battistoni07} and GEANT4~\cite{Agostinelli03} for the fragmentation of primary 400~MeV/nucleon $\noyau{C}{12}{}$ in a thick water target. This work has shown disagreement up to 100\% for the models provided by the GEANT4 toolkit (namely G4BinaryLightIonReaction and G4QMDReaction). Another comparison has been done using the GEANT4 toolkit for a 95~MeV/nucleon $\noyau{C}{12}{}$ on thick PMMA targets \cite{Braunn13}. This study has shown discrepancies up to one order of magnitude as compared to experimental data, especially at forward angles.

In view of this difficulty of the GEANT4 nuclear models to reproduce the fragmentation processes on the energy range useful in carbon-therapy using thick targets, it appeared necessary to constrain these nuclear models with double differential fragmentation cross sections on thin targets. A first set of experimental data has been obtained for a 62~MeV/nucleon $\noyau{C}{12}{}$ on thin carbon target \cite{Napoli12}. GEANT4 simulations results have shown discrepancies up to one order of magnitude for both angular and energy distributions.

A new set of double differential cross section data have been recently obtained by our collaboration (LPC Caen, IPHC Strasbourg, SPhN Saclay, IPN Lyon and GANIL). These data, described in Dudouet {\it et al.} \cite{Dudouet13b} provide good quality measurements (within a 5 to 15\% accuracy) of 95~MeV/nucleon $\noyau{C}{12}{}$ differential cross sections on thin targets (C, CH$_2$, Al, Al$_2$O$_3$, Ti and PMMA). These experimental data are used in this work to test the different nuclear models embedded in the GEANT4 framework. These nuclear models are: G4BinaryLightIonReaction (BIC), G4QMDReaction (QMD) and INCL++. They are coupled to two de-excitation models: the generalized evaporation model (GEM) and the Fermi break-up (FBU). Strengths and weaknesses of these different models in reproducing the fragment production yields, the angular and energy distributions, as well as the target mass dependence will be discussed.


\section{Monte Carlo simulations}\label{subsec:G4Sim}

GEANT4 is a Monte Carlo particle transport code used to simulate the propagation of particles through matter by taking into account both electromagnetic and nuclear processes. It is widely used in a variety of application domains, including medical physics. The 9.6 version of GEANT4 has been used in this work. Electromagnetic interactions are those developed in the ``electromagnetic standard package option 3''. Particle transport cuts have been set to 700 $\mu$m. Total nucleus-nucleus reaction cross sections have been determined, as recommended, using the recently implemented Glauber-Gribov model \cite{Grichine09}. This model provides the full set of nucleus-nucleus cross-sections needed for the GEANT4 tracking (inelastic, elastic, particle production and quasi-elastic) for all incident energies above 100~keV/A.

Nuclear reactions are usually described by a two-step process: a first dynamical step called ``entrance channel'' followed by a de-excitation step called ``exit channel''. The entrance channel model describes the collision and the production of excited nuclear species until thermal equilibrium is achieved. The decay of such hot species is thus considered in a second step by means of statistical de-excitation models. All nuclear models implemented in GEANT4 follow this scheme. In this work, three different entrance channel models are coupled with two exit channel models leading to six different combinations. We stress that the aim of this article is to provide a benchmark of nucleus-nucleus collision models as they are implemented in the GEANT4 toolkit, rather than to test the physical relevance of these models.

\subsection{The GEANT4 entrance channel models}\label{subsubsec:models_in}

Two nuclear models are currently recommended to perform simulations for hadrontherapy. The first one is a binary intra-nuclear cascade (BIC) called G4BinaryLightIonReaction \cite{Wellisch12}. This is an extension of the Binary Cascade model \cite{Folger04} for light ion reactions. This model can be characterized as an hybrid model between a classical cascade code and a quantum molecular dynamics (QMD) description because the 'participating' particles are described by means of gaussian wave functions. By 'participating' particles, it is meant those particles that are either primary particles from the projectile or particles generated and/or scattered during the cascade process. The Hamiltonian is built with a time-independent optical potential. This potential is acting on participants only. Note that in this model, scattering between participants is not taken into account. Participants are tracked until escaping from the nucleus or until the end of the cascade. The cascade stops if the mean kinetic energy of participants in the system is below 15~MeV or if all the participant kinetic energies are below 75~MeV. If such conditions are fulfilled, the system is assumed to have reached thermal equilibrium. The nuclear system is left in an excited state, which evolution toward equilibrium is described by the native pre-compound model of GEANT4.

Another model used in hadron therapy is a QMD-like model called G4QMDReaction \cite{Koi10} adapted from the JAERI QMD (JQMD) code \cite{Niita95,Niita99}. As for the BIC model, the basic assumption of a QMD model is that each nucleon is decribed by a gaussian wave function which is propagated inside the nuclear medium. Differently from the previous model, in the QMD model, all nucleons of the target and of the projectile are taken into account. Each nucleon is thus considered as 'participant'. The particles are propagated and interact by means of a phenomenological nucleon-nucleon potential. The time evolution of the system is stopped at 100~fm/c where it is assumed that equilibrium has been achieved. The QMD model does not include pre-compound model.

A third model has been used in this work: the Li\`ege  Intranuclear Cascade model INCL++ \cite{Boudard13,Kaitaniemi11,Wellisch12}. The last version implemented in GEANT4 is labeled as INCL++ v5.1.8. This model has recently shown promising results \cite{Braunn13} comparable with the BIC or QMD models. Nucleons are modeled as a free Fermi gas in a static potential well. To treat the collision, a target volume is first calculated. Nucleons from the projectile entering this volume are labeled as participants. The quasi-projectile is built from projectile spectators and from non-cascading projectile participants. In contrast, the quasi-target is included in the calculation volume, which also encompasses the participant zone. The final state of the quasi-target is determined by the full collision dynamics of the cascade. Its physical description is therefore much more reliable. The nucleus-nucleus collision is thus not treated symmetrically. Results have shown that INCL better reproduces the target fragmentation than the projectile fragmentation~\cite{Braunn13}. In view of this, INCL treats by default the collision in inverse kinematics (target impinging on projectile), in order to obtain the best reproduction of the projectile fragmentation. However, INCL is not able to use projectile heavier than A=18. If the target is heavier than A=18, the collision will then be performed in direct kinematics. If both target and projectile are heavier than A=18, the description of the collision uses the G4BinaryLightIonReaction model. The effects of this asymmetry in the treatment of the projectile and the target and the discontinuity at mass 18 will be discussed later. The cascade is stopped when no participants are left in the nucleus or when a stopping time defined as : $t_\text{stop} = 70\times(A_\text{target}/208)^{0.16}$~fm/c is reached. As for the G4QMDReaction model, the INCL model does not include pre-compound model.

For the QMD and INCL++ models, a clustering procedure is applied to nucleons. For the QMD model, this clustering procedure is made in phase-space. For the INCL++ model, this clustering procedure is based on a coalescence model. The clustering procedure produces excited species at the end of the cascade. For the BIC model, no clustering procedure is applied and the excitation energies are determined for the projectile and the target remnants. The excitation energy of each species is then estimated and is the input for the de-excitation process considered in the statistical de-excitation codes.

\subsection{The GEANT4 exit channel models}\label{subsubsec:models_out}

GEANT4 provides several de-excitation models which have been recently improved~\cite{Quesada11}. These models describe particle evaporation from excited nuclear species produced in the entrance channel. Two models have been considered in this work.

The first one is the generalized evaporation model (GEM)~\cite{Furihata00,Wellisch12}. Based on the Weisskopf-Ewing evaporation model~\cite{Weisskopf37}, it considers sequential particle emission up to $\noyau{Mg}{28}{}$ as well as fission and gamma decay.

The second model is the Fermi Break-up model (FBU)~\cite{Wellisch12}. This model considers the decay of an excited nucleus into $n$ stable fragments produced in their ground state or in low-lying discrete states. The break-up probabilities for each decay channel are first calculated by considering the n-body phase space distribution. Such probabilities are then used to sample the decay channels by a Monte-Carlo procedure. This model is only used for light nuclei (Z$\leq$8 and A$\leq$16). For heavier nuclei, the de-excitation process is considered
using the GEM model.

\section{Experimental data}\label{subsec:ExpData}

The models described above will be compared with data obtained during the E600 experiment performed in May 2011 at the GANIL facility (Grand Acc\'el\'erateur National d'Ions Lourds). The experiment has allowed to measure the double differential cross sections of various species in 95~MeV/nucleon $\noyau{C}{12}{}$ reactions on H, C, O, Al and $^{nat}$Ti targets~\cite{Dudouet13b}. The description of the experimental set-up and the experimental energy thresholds are described in Dudouet {\it et al.}~\cite{Dudouet13b}.The particles have been detected using three stages telescopes, located at angles ranging from 4$^\circ$ to 43$^\circ$. They have been identified using a $\Delta$E-E method. The analysis method has been described in Dudouet {\it et al.}~\cite{Dudouet13a}. The errors bars of the presented experimental data are including systematics and statistical errors. These data are available with free access on the following web-site \url{http://hadrontherapy-data.in2p3.fr}.

\begin{table}[!ht]
\begin{center}
{\setlength{\tabcolsep}{0.8mm}
{\renewcommand{\arraystretch}{1.5}
\begin{tabular}{|c|c|c|c|c|c|c|c|c|c|}
\cline{1-9}
isotope&$^1$H&$^2$H&$^3$H&$^3$He&$^4$He&$^6$He&$^6$Li&$^7$Li \\	
\cline{1-9}
E$_{\text{th}}$ (MeV)&4.0&5.2&6.1&14.2&16.0&18.6&29.9&31.7\\	
\cline{1-9}
\multicolumn{10}{c}{\vspace{-5mm}}\\
\hline
isotope&$^7$Be&$^9$Be&$^{10}$Be&$^8$B&$^{10}$B&$^{11}$B&$^{10}$C&$^{11}$C&$^{12}$C \\	
\hline
E$_{\text{th}}$ (MeV)&44.3&48.6&50.5&60.6&65.8&68.1&81.3&84.2&86.9\\	
\hline
\end{tabular}}}
\caption{Energy threshold used in the simulations.}
\label{table:seuils}
\end{center}
\end{table}

In the presented simulations, only the energy thresholds of the telescopes have been taken into account. As a reminder, these thresholds are shown in table \ref{table:seuils} for all the detected isotopes. The main effects of these thresholds is to lower the contribution of the particles coming from the target fragmentation. It has been verified that the presented simulations and the simulations in which the whole set-up is taken into account give the sames results. The main drawback of these latter simulations is their lack of CPU efficiency since the fragmentation process in thin targets is rare and the solid angles of detectors are small. The target thicknesses used in the simulations are the same than the experimental one in order to obtain the same angular and energy straggling. A number of $10^{9}$ incident $\noyau{C}{12}{}$ has been used in the simulations in order to minimize the statistical error on the simulated data (lower than 1\% in most cases but up to 20\% for the larger angles). The choice has been made to not represent the statistical errors of the simulated data for clarity reasons.

\section{Results}\label{sec:Results}

\subsection{The participant-spectator scheme}\label{subsec:reac_mech_rep}

\begin{figure*}[!ht]
\begin{center}
\includegraphics[width=0.9\linewidth]{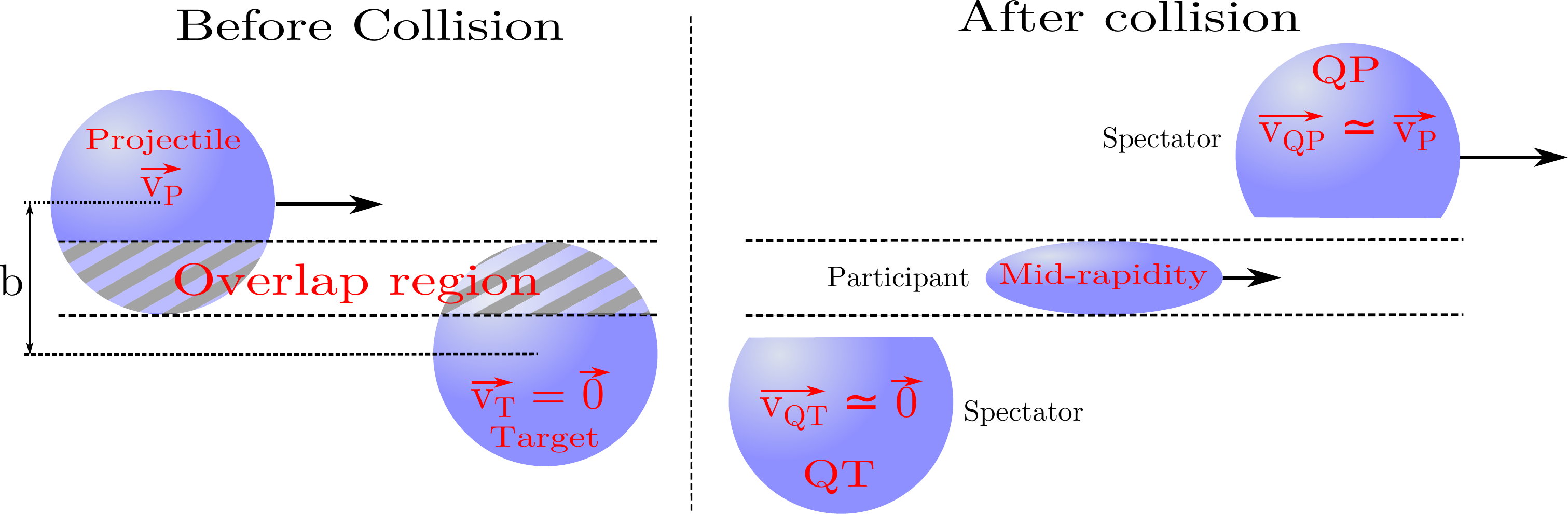}
\caption{(Color online) Schematic representation of the geometrical participant-spectator model in the laboratory frame.}
\label{Fig1}
\end{center}
\end{figure*}

Some characteristics of the results will be discussed in the framework of the participant-spectator picture of the collision (see for instance Fig.~\ref{Fig1}~\cite{Durand01,Babinet85}).  

This is a typical high energy process (in the GeV/A range) in which the internal velocities of the nucleons are (much) smaller than the relative velocity between the two partners of the reactions. However, recent analysis have shown that it could still be valid around 100~MeV/nucleon incident energy \cite{Dudouet13b}. In such a picture, for a finite impact parameter, b, the nucleons located in real space in the overlapping region of the two nuclei constitute the 'participants'. The projectile nucleons outside the overlapping region constitute the moderately excited quasi-projectile moving with a velocity close to the beam velocity. The same argument applies for the target nucleons leading to a quasi-target moving with a velocity close to 0. The participants constitute the so-called highly excited mid-rapidity source. The decay products from this source show an energy distribution shifted towards lower values as compared to the beam energy. Therefore, in such a picture, three energy contributions in the laboratory frame are expected: a first one close to the beam energy, a second one associated with the target at energies close to 0 and in between, a contribution associated with the participants. This latter is thus to a large extent strongly coupled to the sizes of the projectile and of the target and should show up as the size of the target increases. We stress that this very simple picture is used here to define the terms that characterize the origins of the detected fragments and not used as a realistic description of the reaction mechanisms. 

The results of the models considered above are now compared with experimental data. We first consider a comparison of simulated  cross sections (production, angular and energy distributions) with the experimental data in the case of carbon target. Then, the target mass dependence will be studied.

\subsection{Production cross sections}\label{subsec:ProdRates}

\begin{figure*}[!ht]
\includegraphics[width=1\linewidth]{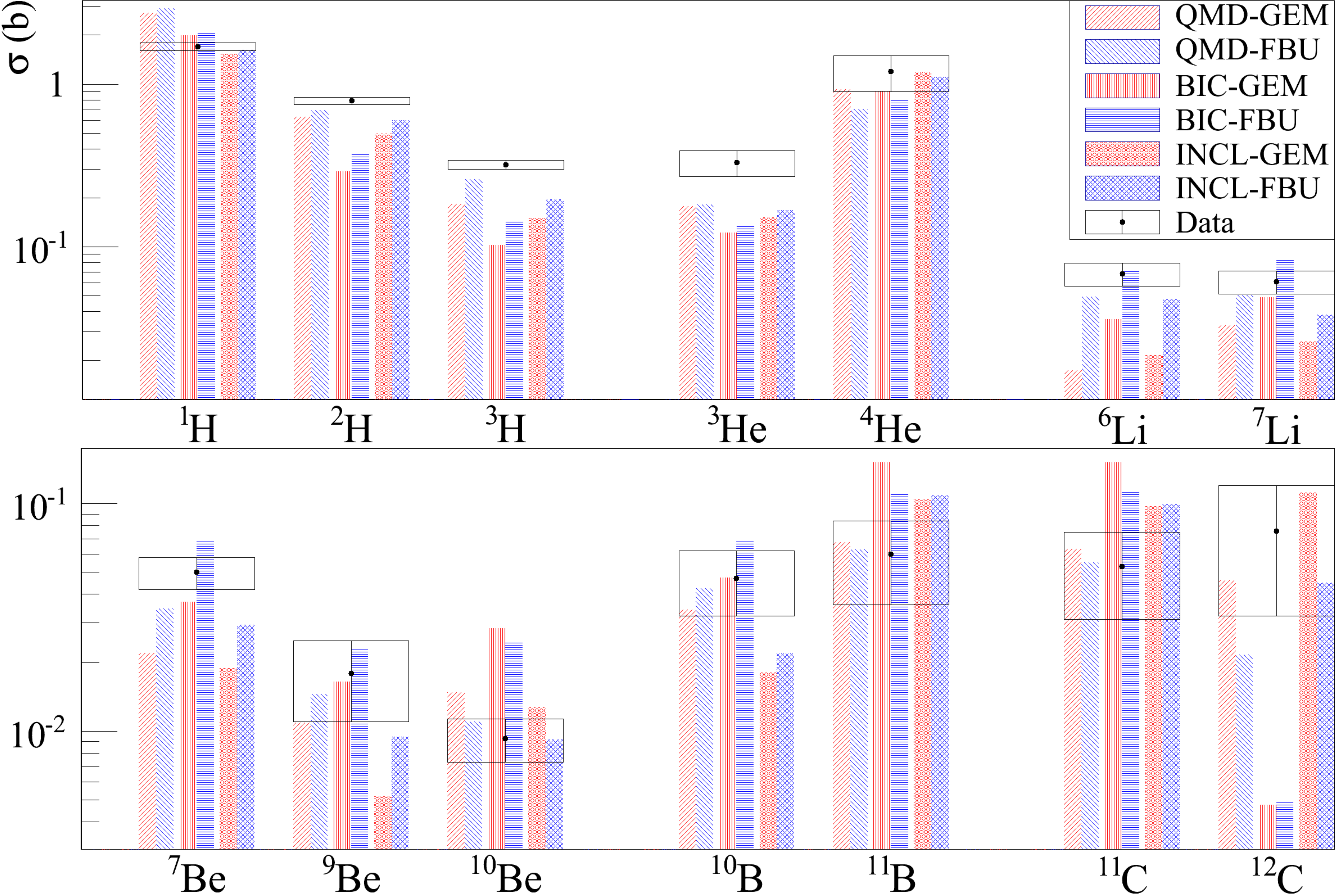}
\caption{(Color online) Comparisons between data and the different combination of entrance and exit channel models (see text) for the production cross sections of various isotopes in  95~MeV/nucleon~$\noyau{C}{12}{}\rightarrow\noyau{C}{12}{}$ reactions.}
\label{Fig2}
\end{figure*}

Fig.~\ref{Fig2} displays the production cross sections of the most abundant reaction products in the case of a carbon target. They are compared with the GEANT4 results with the different combinations between the entrance and exit channel models discussed previously. Note that the production cross section of $\noyau{C}{12}{}$ fragments takes into account only inelastic interactions, excluding elastic scattering. These production cross sections have been obtained by fitting the angular distributions with a function resulting of the sum of a gaussian and an exponential function. These fitted function have then been integrated over the whole solid angle~\cite{Dudouet13b}. The errors bars represented on Fig.~\ref{Fig2} have been obtained by propagating the fit parameters uncertainties, using the covariance matrix of the fit procedure.

The results of Fig.~\ref{Fig2} clearly shows that none of the model combination is able to reproduce the production rates for all isotopes. Moreover, it is not easy to identify which model combination is the most suited for a comparison with experimental data. However, it may be concluded that the influence of the entrance channel is larger than the influence of the exit channel model. Regarding the two exit channel models, the Fermi Break-up model seems, for a given entrance channel model, to be, in most cases, more compatible with the data. This was already mentioned in B\"ohlen {\it et al.}~\cite{Bohlen10} and Ivanchenko {\it et al.}~\cite{Ivanchenko12}. This is due, to some extent, to the fact that the Fermi Break-up description allows to explore more available phase space (especially at high excitation energies for which three (or more) body decay may play an increasing role) than the GEM model for which only sequential evaporation is taken into account. In the following, we only consider calculations in which the Fermi Break-up model is used for the exit channel.

\subsection{Angular distributions}\label{subsec:AngDist}

\begin{figure*}[!ht]
\includegraphics[width=\linewidth]{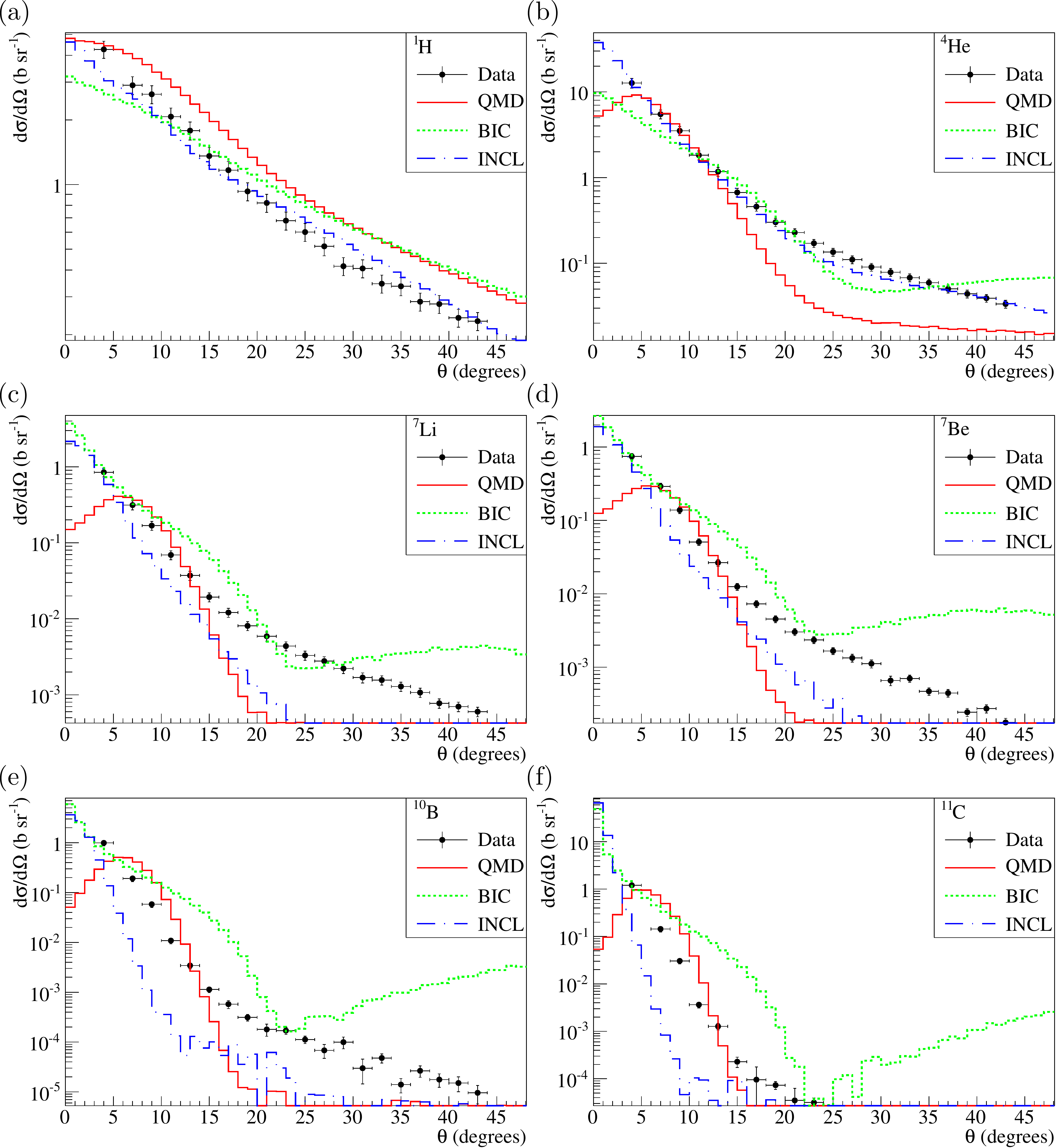}
\caption{(Color online) Absolute differential angular cross-sections of protons, $\noyau{He}{4}{}$, $\noyau{Li}{6}{}$, $\noyau{Be}{7}{}$, $\noyau{B}{10}{}$ and $\noyau{C}{11}{}$ obtained for the carbon target. Experimental data: black points. Histograms: GEANT4 simulations with QMD, BIC and INCL models coupled to the Fermi Break-up de-excitation model as indicated in the insert.}
\label{Fig3}
\end{figure*}

The E600 experimental setup allowed to cover an angular range from 4$^\circ$ to 43$^\circ$ by steps of two degrees. Fig.~\ref{Fig3} displays the differential angular cross-sections for carbon target for both experimental data and for simulations using QMD, BIC, and INCL models coupled with the Fermi Break-up de-excitation model.

Although QMD is the most achieved model as far as the dynamics of the collision is concerned, it fails to reproduce the angular distributions. It strongly overestimates the proton production of about 50\% (as also observed in Fig.~\ref{Fig2}), and poorly reproduces the angular distributions of the heavier isotopes considered here (up to one order of magnitude). The distibutions obtained with QMD show maximum values around 7$^\circ$ (apart for protons) with a fall off towards $0^\circ$. This is in disagreement with the experimental distributions showing an increase at very low angles.

The distributions obtained with the BIC model are slightly closer to the data as compared with QMD, especially at forward angles and for heavier fragment distributions ($\noyau{Li}{6}{}$ and $\noyau{Be}{7}{}$). The lack of $\alpha$ at forward angles may possibly come from a failure of the model to take into account the $^{12}$C three alpha cluster structure. The global shape is however not correct. The quasi-projectile contribution is too large and the large angles are poorly reproduced. The angular distributions obtained with the BIC model increases around 25$^\circ$ (except for protons). This probably comes from the quasi-target contribution but is in disagreement with experimental data.

Finally, INCL is the model that seems to better reproduce the angular distributions, especially for light fragments. The shapes of protons and $\alpha$ distributions are nearly reproduced over the whole angular range ($\sim$ 10-20\%), despite a small underestimation of the protons at forward angles. Regarding the distributions of heavier fragments, as for the BIC model, only the forward angles are well described. At large angles the INCL model strongly underestimate the data (up to one order of magnitude).  

We have shown in Dudouet {\it et al.}~\cite{Dudouet13b} that the experimental angular distributions of particles emitted in the 95~MeV/nucleon $\noyau{C}{12}{}$ reaction on H, C, O, Al and $^{nat}$Ti can be represented as the sum of a gaussian and an exponential contribution. None of the models used here are able to reproduce this trend. The main problem is associated with the inability of such models to reproduce the magnitude of the exponential contribution which is dominant at large angles. Since this contribution is mostly resulting from the mid-rapidity source discussed previously, it is tempting to conclude at this stage that the present models do not contain the ingredients needed to describe the mid-rapidity processes. We now proceed with the energy distributions.

\subsection{Energy distributions}\label{subsec:EnerDist}

\begin{figure*}[!ht]
\includegraphics[width=\linewidth]{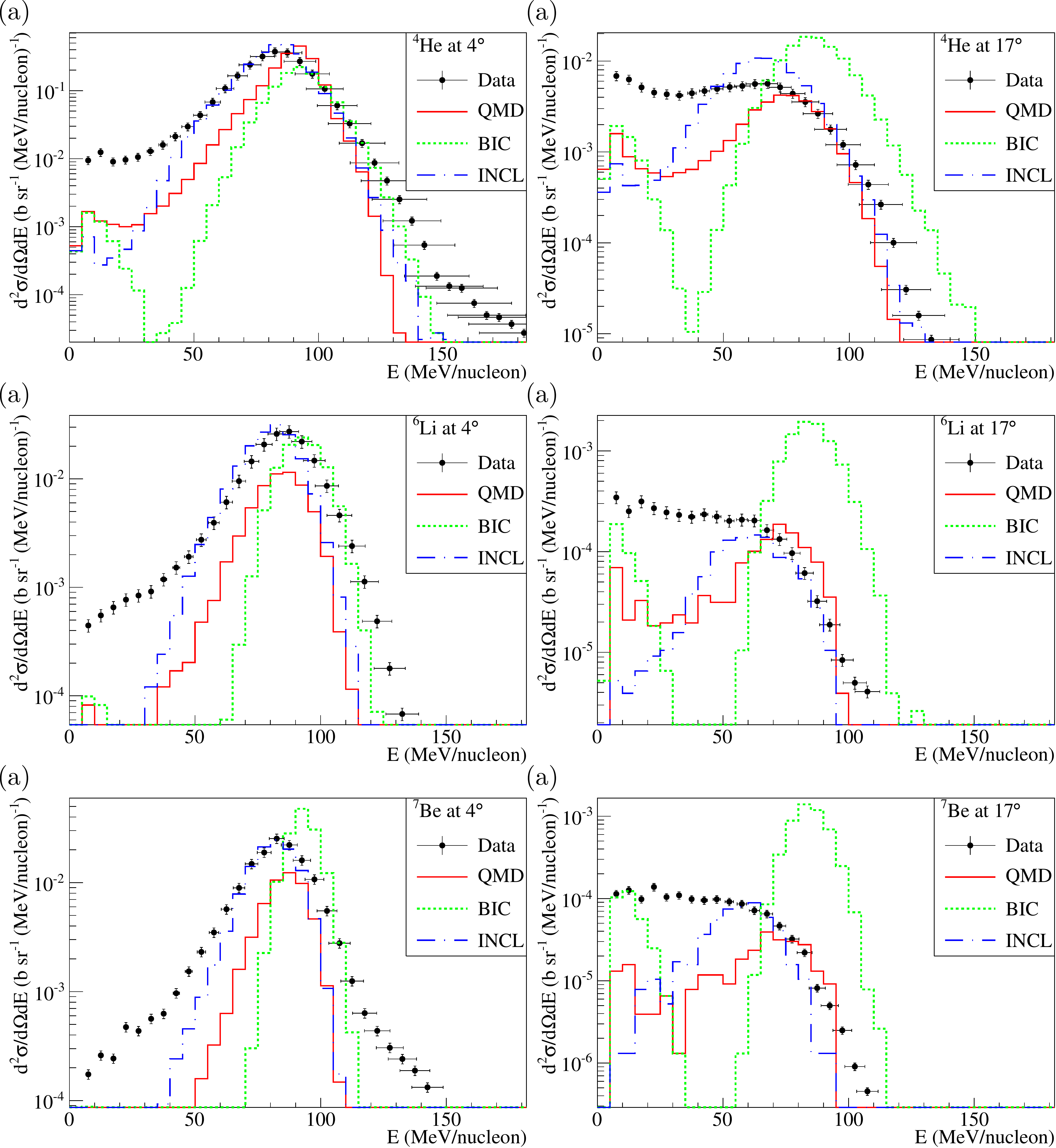}
\caption{(Color online) Energy distributions of $\noyau{He}{4}{}$, $\noyau{Li}{6}{}$ and $\noyau{Be}{7}{}$ fragments at 4 and 17$^\circ$. Black points: experimental data. Histograms are for simulations with QMD, BIC and INCL models coupled to the Fermi Break-up de-excitation model (see insert).}
\label{Fig4}
\end{figure*}

The agreement with the double differential cross sections constitutes the most severe test of the models. Fig.~\ref{Fig4} shows few examples of energy distributions obtained for $\noyau{He}{4}{}$, $\noyau{Li}{6}{}$ and $\noyau{Be}{7}{}$ at 4 and 17$^\circ$.

Here, we would like to focus on the shape of the distributions rather than on the absolute magnitude. The distributions may be interpreted as follows: the major contribution originates from the decay of the quasi-projectile and is thus located at an energy close to the beam energy per nucleon. This peak close to the beam energy is clearly visible at small angles (cf Fig.~\ref{Fig4}: distibutions at 4$^\circ$). At larger angles, this contribution tends to vanish because of the strong focusing of the quasi-projectile. The low energy part of the distribution is associated with the species produced at mid-rapidity and also with the decay of the quasi-target although this last contribution becomes dominant only at very large angles and can be poorly detected due to experimental energy thresholds. Therefore, the ability of the models to reproduce the data can be appreciated on these two physical aspects: the decay of the projectile-like and the particle production mechanism at mid-rapidity. 

As shown in Fig.~\ref{Fig4}, among the three models, BIC shows the strongest disagreement with the experimental data. In particular, the model is unable to account for the mid-rapidity contribution (medium angles). This is due to the binary nature of the reaction mechanism assumed in the model. Indeed, composite fragments cannot be formed in this model and only nucleons undergoing nucleon-nucleon collisions can be emitted. Moreover, the mean energy of the quasi-projectile contribution is too large as compared to data and its contribution (close to 95~MeV/u) remains too important at large angle. This leads for instance to the very strong disagreement shown in Fig.~\ref{Fig4} (d) for $\noyau{Be}{7}{}$ fragments at 17$^\circ$.

The INCL model better reproduces  the quasi projectile contribution both for the mean and the width of the energy distribution. It also predicts more fragments at low energies ($0<$E$<50$ MeV/nucleon) as compared to the BIC model. However, the results still underestimate the data. Moreover, the shape of the distributions at low energies (mid-rapidity contribution) is not in agreement with the data.

Contrary to angular distributions, the QMD model better reproduces the global shape of the energy distributions. Although the mean energy of the quasi-projectile peak is slightly too high, the shape of the mid-rapidity contribution is better reproduced than for the BIC or INCL models. However, as for other models, it underestimates the mid-rapidity contribution.

The remarks mentioned above are valid for all fragments from protons to carbon isotopes. The main conclusion that can be drawn is that none of the tested models is able to reproduce simultaneously the quasi-projectile, the quasi-target and the mid-rapidity contributions. The INCL model better reproduces the quasi-projectile contribution: it is probably the best model for the description of the quasi-projectile. In contrast, the QMD model better describes the mid-rapidity emission, probably due to the fact that it is the only model to take into account the time propagation and the interaction of all the nucleons in the reaction. Similar conclusions have been drawn at lower energy in De Napoli {\it et al.}~\cite{Napoli12}, where the BIC and QMD models were tested in 62~MeV/nucleon~$\noyau{C}{12}{}\rightarrow\noyau{C}{12}{}$ induced reactions.

\subsection{Results with other targets}\label{subsec:TarComp}

\begin{figure*}[!ht]
\includegraphics[width=\linewidth]{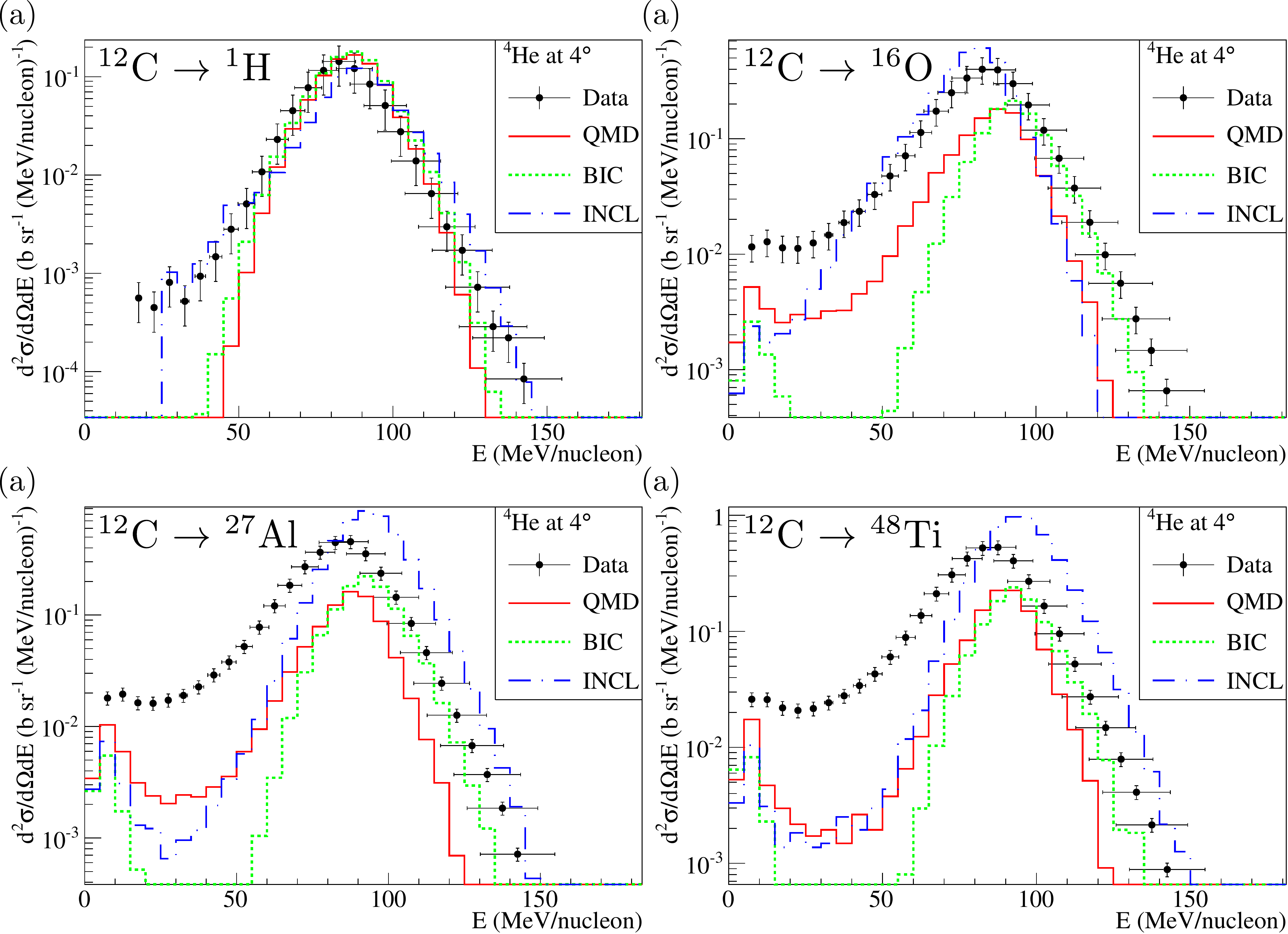}
\caption{(Color online) Energy distributions of $\alpha$ particles at 4$^\circ$ for hydrogen (a), oxygen (b), aluminum (c) and titanium (d) targets. Black points: experimental data. Histograms: simulations (see insert).}
\label{Fig5}
\end{figure*}

Our experiment allowed to gather data for a series of targets ranging from hydrogen up to titanium. The target dependence on the double differential cross sections is now investigated. Fig.~\ref{Fig5} displays the $\alpha$ energy distributions at 4$^\circ$ for the hydrogen, oxygen, aluminum and titanium targets for both data and simulations using QMD, BIC and INCL models coupled to the Fermi Break-up de-excitation model.

The three models reproduce quite well the data for the hydrogen target, especially INCL. This result is not surprising in the sense that these models are mostly based on the concept of nuclear cascade which was originally dedicated to nucleon-nucleus collisions. In such reactions, the geometry of the collision is rather simple and the description of the quasi-projectile is easier than for nucleus-nucleus reactions. More, with the hydrogen target, the $\alpha$ particles are mainly produced by the quasi-projectile de-excitation. However, the experimental data exhibits a small contribution at low energy (below 50~MeV/nucleon) and INCL is the only model to reproduce this contribution. 

Nevertheless, the heavier the target, the larger the disagreement between the simulations and the experimental data. From carbon to titanium, the three models reproduce quite well the quasi-target and the quasi-projectile contributions. The difficulty to produce mid-rapidity fragments is evidenced. The discrepancy is amplified as the target mass increases emphasizing the increasing role of mid-rapidity in the data as a simple consequence of the geometry of the reaction. The larger the mass of the target, the larger the size of the mid-rapidity region. The BIC model does not produce mid-rapidity fragments (around E = 40-50~MeV/nucleon). Although the situation is slightly better for INCL or QMD models, the mid-rapidity contribution is underestimated for both models.

A particular attention needs to be paid to the INCL model. For the aluminum and titanium targets, the shape of the energy distribution changes with respect to lighter targets. The projectile contribution is overestimated and the mean energy is too large. The reason is due to the discontinuity in the treatment of the kinematics when the target is larger than A=18 as mentioned in Sec.~\ref{subsubsec:models_in}. Otherwise, for lighter targets, results concerning the quasi-projectile are promising while the production at mid-rapidity remains underestimated.

In the participant-spectator reaction mechanism, the mid-rapidity contribution originates from the overlap region as already mentioned previously. This is thus a geometrical contribution, which increases significantly with the target size, as it is observed experimentally when going from the hydrogen to the titanium target: more and more fragments are produced in the low energy region. The three models that have been used here fail in accurately reproducing this region and the discrepancy increases with the mass of the target. This may be due to the fact that none of them take accurately into account the possibility to produce sizeable clusters in the overlapping region. This point should deserve additional studies.

\section{Conclusions}\label{sec:Conclusions}

In this work, comparisons have been performed between experimental data collected in 95~MeV/nucleon $\noyau{C}{12}{}$ reactions on H, C, O, Al and $^{nat}$Ti targets and GEANT4 simulations in order to test the models embedded in the GEANT4 nuclear reaction package. The G4BinaryLightIonReaction (BIC), the G4QMDReaction (QMD) and the INCL++ (INCL) entrance channel models have been coupled to the generalized evaporation model (GEM) and the Fermi break-up model (FBU) exit channel models.

The main conclusion is that up to now, none of these six models combinations is able to accurately reproduce the data, neither in term of production rates nor for angular or energy distributions.

This study has shown that the entrance channel model characteristics have a larger effect on particles and fragments production as compared to the choice of the exit channel description. However, the Fermi break-up de-excitation model seems to give better results than the generalized evaporation model. This observation has also been done in B\"ohlen {\it et al.}~\cite{Bohlen10} and Ivanchenko {\it et al.}~\cite{Ivanchenko12}.

As for angular distributions, apart from INCL which reproduces quite well protons (with a small disagreement at forward angles) and $\alpha$ distributions for the carbon target, the models are not able to reproduce the data. The QMD model is the worst, with a maximum value of the distribution at around 7$^\circ$ and an unexpected fall off towards 0$^\circ$.

On the contrary, QMD is the one which better reproduces the energy distributions for all considered fragments. Apart from the hydrogen target, the BIC model fails to reproduce the data and in particular, it does not produce particles at low energy. The INCL model reproduces very well the quasi-projectile contribution if the target is not larger than A=18.

These results seems consistent with those observed at lower energy. Indeed, the GEANT4 simulations that have been done in De Napoli {\it et al.}~\cite{Napoli12} have shown that the angular distributions were better reproduced by the BIC model than the QMD model. Regarding the energy distributions, it has been shown that the QMD model better reproduces the shape of the distribution than the BIC one. The conclusions on the GEANT4 nuclear models that we have made at 95~MeV/nucleon are thus in agreement with the one made at lower energies. The better reproduction of carbon fragmentation processes for the QMD model than for the BIC one has also been observed on thick water target at higher energies in B\"ohlen {\it et al.}~\cite{Bohlen10}. However, no INCL simulations have been performed in these two studies.

Finally, a study of the target mass dependence shows that the three models do not succeed in reproducing realistically  the production of species at mid-rapidity. Comparisons with a simple phenomenological model that takes into account the geometrical overlap region is planned in a near future.

\bibliographystyle{unsrt}
\bibliography{G4_article}

\end{document}